\documentclass[12pt]{article}
\usepackage[a4paper]{geometry}
\usepackage[italian,english]{babel}

\usepackage{amsmath}
\usepackage{amsfonts}
\usepackage{amstext}
\usepackage{amssymb}
\usepackage{amsthm}
\usepackage{amscd}

\usepackage[pagebackref,draft=false]{hyperref}
\hypersetup{colorlinks,
linkcolor=myrefcolor,
citecolor=mycitecolor,
urlcolor=myurlcolor}

\usepackage[capitalize]{cleveref}
\usepackage{caption}
\usepackage{etaremune}

\usepackage{xcolor}
\definecolor{myurlcolor}{rgb}{0,0,0.4}
\definecolor{mycitecolor}{rgb}{0,0.5,0}
\definecolor{myrefcolor}{rgb}{0.5,0,0}
\usepackage{graphicx}
\usepackage{tikz}
\usepackage{tikz-cd}
\usepackage{mathrsfs}

\usepackage{etoolbox}
\usepackage{makeidx}
\usepackage{sectsty}
\usepackage{dsfont}
\usepackage{enumitem} 
\usepackage[]{latexsym}
\usepackage{braket}
\usepackage{caption}
\usepackage[utf8]{inputenx}
\usepackage[T1]{fontenc}
\usepackage{lmodern}
\usepackage{textcomp}
\usepackage{microtype}
\usepackage{totcount}
\usepackage{blindtext}
\newtheorem*{proof*}{Proof}

\newcommand{\be}{\begin{equation}}
\newcommand{\ee}{\end{equation}}
\newcommand{\bea}{\begin{eqnarray}}
\newcommand{\eea}{\end{eqnarray}}

\newcommand{\grit}[1]{{\bfseries {\itshape {#1}}}}

\newcommand{\lra}{\longrightarrow}

\newcommand{\hh}{\mathcal{H}}

\newcommand{\stsp}{\mathscr{S}}

\newcommand{\appa}{\mathscr{A}}

\newcommand{\V}{\mathscr{V}}

\title{On the notion of composite system}

\author{F. M. Ciaglia$^{1,6}$\href{https://orcid.org/0000-0002-8987-1181}{\includegraphics[scale=0.7]{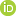}}, A. Ibort$^{2,3,7}$\href{https://orcid.org/0000-0002-0580-5858}{\includegraphics[scale=0.7]{ORCID.png}},  G. Marmo$^{4,5,8}$\href{https://orcid.org/0000-0003-2662-2193}{\includegraphics[scale=0.7]{ORCID.png}} \\
\footnotesize{$^{1}$\textit{ Max Planck Institute for Mathematics in the Sciences, Leipzig, Germany}} \\
\footnotesize{$^{2}$\textit{ ICMAT, Instituto de Ciencias Matem\'{a}ticas (CSIC-UAM-UC3M-UCM)}}  \\
\footnotesize{$^{3}$\textit{ Depto. de Matem\'aticas, Univ. Carlos III de Madrid, Legan\'es, Madrid, Spain}}  \\
\footnotesize{$^{4}$\textit{ Dipartimento di Fisica ``E. Pancini'', Universit\`a di Napoli Federico II, Napoli, Italy}} \\
\footnotesize{$^{5}$\textit{ INFN-Sezione di Napoli, Napoli, Italy.}} \\
\footnotesize{$^{6}$\textit{ e-mail: \texttt{florio.m.ciaglia[at]gmail.com}}, $^{7}$\textit{ e-mail: \texttt{albertoi[at]math.uc3m.es}}} \\
\footnotesize{$^{8}$\textit{ e-mail: \texttt{marmo[at]na.infn.it}}}
}

\date{}

\begin{document}

\maketitle

\begin{abstract}
The notion of composite system made up of distinguishable parties is investigated in the context of arbitrary convex spaces.
\end{abstract}

\section{Introduction}

When dealing with composite systems, one of the most striking features of quantum theories is undoubtedly the existence of non-classical correlations between subsystems of the given system, a phenomenon known under the name of `entanglement'.

In the context of standard quantum mechanics \cite{dirac-principles_of_quantum_mechanics,pauli-general_principles_of_quantum_mechanics,von_neumann-mathematical_foundations_of_quantum_mechanics}, where a physical system is described by means of a Hilbert space $\hh$ and the physical states are density operators on $\hh$, entanglement is associated with the fact that the Hilbert space of a composite system is not the Cartesian product of the Hilbert spaces of the subsystems as it happens for the phase space of a classical composite system, but, rather, it is taken to be the tensor product of the Hilbert spaces of the subsystems.

A more refined formalism for quantum theories is the algebraic formulation in terms of $C^{*}$-algebras \cite{araki-mathematical_theory_of_quantum_fields,haag-local_quantum_physics,haag_kastler-an_algebraic_approach_to_quantum_field_theory,jordan_von_wigner-on_an_algebraic_generalization_of_the_quantum_mechanical_formalism,segal-postulates_for_general_quantum_mechanics,segal-c*algebras_and_quantization}.
In this context, a physical system is described in terms of the $C^{*}$-algebra $\appa$ of (bounded) observables  and the physical states are the mathematical states on $\appa$, that is, the positive linear functionals on $\appa$ normalized to $1$.

The reformulation of quantum theories in terms of $C^{*}$-algebras also helps to clearly see the link between quantum theories and classical probability theory.
Indeed, the space of quantum states and the space of probability distributions on a topological/measure space may be described by means of the ``same object'', namely the space of mathematical states on a $C^{*}$-algebra.
When this algebra is Abelian (commutative) we obtain the case of classical probability theory, while when the algebra is non-Abelian, we enter in the quantum realm.
Analogously to what happens in the Hilbert-space formalism of quantum theories, the entanglement content of the theory comes from the fact that the $C^{*}$-algebra of a composite system is  taken to be a suitable tensor product of the $C^{*}$-algebras of some subsystems.

In this contribution,  we want to understand the mathematical requirements we should impose on the description of the notion of composite system in a given theoretical framework in order for the tensor product of suitable objects to necessarily come out.
What we have in mind is a rather elementary discussion on the mathematical features characterizing the relation between composite systems and tensor products.
Accordingly, in order to avoid as much as we can to rely on the specific traits of some given theoretical framework, we will not focus much on the technical and interpretational details.

Essentially, we will model the space of states of a physical system  by means of a real, convex space $\stsp$.
This is a very broad theoretical framework of which the spaces of states of both classical probability theory and quantum theories are a particular instance.
From the operational point of view, this perspective is motivated by the idea that the states of a physical system are associated with equivalence classes of preparation procedures yielding the same measurement statistics, and that inequivalent preparation procedures may be ``mixed in arbitrary proportions''  resulting in operations that may be considered as admissible preparation procedures (see \cite{cornette_gudder-the_mixture_of_quantum_states,gudder-convex_structures_and_operational_quantum_mechanics,hellwig_kraus-pure_operations_and_measurements,holevo-statistical_structure_of_quantum_theory,holevo-probabilistic_and_statistical_aspects_of_quantum_theory,kraus-states_effects_and_operations,ludwig-foundations_of_quantum_mechanics_I,mielnik-geomery_of_quantum_states,mielnik-theory_of_filters,mielnik-generalized_quantum_mechanics}).
Mathematically speaking, this instance is then translated in the possibility of taking arbitrary convex combinations of elements in $\stsp$ describing physical states.

From the purely mathematical point of view, the fact that $\stsp$ is a convex set implies the existence of the vector space $\stsp^{*}$ of real-valued, affine linear functionals on $\stsp$, and this space will be the only ingredient, beside $\stsp$, we will need in our discussion.
Note that $\stsp^{*}$ coincides with the dual space $\V^{*}$ of the vector space $\V$ generated by formal linear combinations of elements in $\stsp$.
For the sake of linguistic simplicity, we \grit{define} $\stsp^{*}$ to be the \grit{dual space} of the convex set $\stsp$ with an evident abuse of nomenclature.

We want to stress that, by focusing only on the convex structure of $\stsp$ and its dual space $\stsp^{*}$, our analysis clearly applies to both the space of quantum states and the space of classical probability distributions on a topological/measure space, while mantaining open the possibility of considering different types of theories like those considered in the so-called \grit{generalized probabilistic theories} (see \cite{barnum_wilce-postclassical_probability_theory,barrett-information_processing_in_generalized_probabilistic_theories,chiribella_dariano_perinotti-probabilistic_theories_with_purification,janotta_hinrichsen-generalized_probability_theories:what_determines_the_structure_of_quantum_theory}).

\section{Composite systems and tensor products}

When describing composite systems from a theoretical point of view, there are, essentially, two possible perspectives: either we start from the total system and then proceed in determining a suitable notion of subsystem, or we start from the subsystems and then proceed in determining a suitable notion of composite system.
Here, we will investigate the latter case  in the context of composite systems made of distinguishable parties (we refer to \cite{chiribella-agents_subsystems_and_the_conservation_of_information} for a modern approach to the former case).

For the purpose of this contribution, we represent a composite system by means of the family $\{\stsp_{j}\}_{j\in[1,...n]}$ of spaces of states of the $n$ subsystems together with the space $\stsp$ of states of the total system.
As said before, we consider the spaces of states of the subsystems as given, and we want to characterize the admissible candidates for the convex set of the total system on the basis of additional constraints associated with the notion of composite system.
We shall not deal with indistinguishable ``particles'', i.e., neither Bosons, Fermions or other parastatistics.
These additionial aspects would only add complications without helping in addressing the core problem.
If needed, we can include other types of ``statistics'' at later time.

First of all, we want to implement a notion of ``independence'' among the states (preparation procedures) of the subsystems.
Roughly speaking, we want to formalize the idea according to which there are no constraints among the preparation procedures of the subsystems, that is, each party is free to prepare its associated subsystem in any of the possible states independently from the preparations of the other parties.
Mathematically speaking, we implement this idea by assuming the existence of an injective map 
\be\label{eqn: independence of preparation procedures}
\mathrm{I}\colon \stsp_{1}\times\cdots\times\stsp_{n}\,\lra\,\stsp
\ee
so that for every $n$-tuple $(\rho_{1},...,\rho_{n})\in\stsp_{1}\times\cdots\times\stsp_{n}$ of states there is a corresponding $\rho\in\stsp$ representing the $n$-tuple of independent states (preparation procedures) as a state of the total system.
The notion of independence among the states of the subsystems (see equation \eqref{eqn: independence of preparation procedures}) appears also in the context of algebraic quantum field theory.
For instance, in \cite{roos-independence_of_local_algebras_in_quantum_field_theory}, this condition, together with a commutativity assumption, is used  to prove that the algebra $\mathfrak{A}$ generated by the algebras $\mathfrak{A}_{1}$ and $\mathfrak{A}_{2}$ of observables associated with two space-like separated spacetime regions is isomorphic with the algebraic tensor product $\mathfrak{A}_{1}\otimes\mathfrak{A}_{2}$.
Here, we will obtain a similar result in the framework of convex spaces (of which the spaces of states of $C^*$-algebras typical of algebraic quantum field theory form a subfamily) by replacing the commutativity assumption with an interdependence condition among the dual spaces of the subsystems (see below).

Before proceeding further, we note that the choice $\stsp =\stsp_{1}\times\cdots\times\stsp_{n}$, where $\stsp$ is endowed with the convex sum obtained by the component-wise application of the convex sums of the $\stsp_{j}$'s, is clearly the minimal choice compatible with the assumption of independence among the states  of the subsystems.
In this case, denoting by $\V_{j}$ the vector space canonically generated by $\stsp_{j}$ by means of formal linear combinations of elements in $\stsp_{j}$, it is clear that $\stsp=\stsp_{1}\times\cdots\times\stsp_{n}$ is a subset of the vector space $\V=\oplus_{j=1}^{n}\V_{j}$.
Consequently, the dual space $\stsp^{*}$ of $\stsp= \stsp_{1}\times\cdots\times\stsp_{n}$ is just the dual space of $\V$, that is,  $\V^{*}=\oplus_{j=1}^{n}\V_{j}^{*}$.
In particular, this means that, for every $n$-tuple $(f_{a_{1}},\cdots,f_{a_{n}})$ with $f_{a_{j}}\in\stsp_{j}^{*}$ for $j\in[1,...,n]$, there is an element $f_{a_{1},...,a_{n}}\in\stsp^{*}$  such that
\be\label{eqn: independent dual spaces}
f_{a_{1},...,a_{n}}(\rho_{1},\,\cdots,\,\rho_{n})\,=\,\sum_{j=1}^{n}\,f_{a_{j}}(\rho_{j})\,.
\ee
Clearly,  $f_{a_{1},...,a_{n}}$ vanishes on the product space $\stsp_{1}\times\cdots\times\stsp_{n}$ representing independent equivalence classes of preparation procedures if and only if $f_{a_{j}}=0$ for all $j\in[1,...,n]$.
Intuitively speaking, we may say that the dual spaces of the subsystems do not ``compose'' with each other.
This means that the system described by $\stsp=\stsp_{1}\times\cdots\times\stsp_{n}$ endowed with the component-wise convex sum should be interpreted more as a juxtaposition  rather than a composition of systems, and, in general, we want to avoid the possibility of this convex set as an admissible space of states.

At this point we may say that this is exactly what happens in the groupoid interpretation of Quantum Mechanics \cite{CIM-Pedagogical-Schwinger,CIM-Schwinger,CIM2-Schwinger}. There are two natural operations with groupoids: disjoint union and direct product. The first corresponds to juxtaposition (the corresponding algebra and space of states are direct sums) and the second is the  proper  composition (tensor product).

A possible way to overcome this instance and force the subsystems to   ``compose'' is to implement a notion of interdependence for the dual spaces of the subsystems.
Before introducing this notion of interdependence, we want to point out that there is no clear and unambiguous physical interpretation for it  at this moment because the theoretical framework of  arbitrary  convex spaces does not allow a clear and unambiguous physical interpretation for the dual spaces.
Having cleared this point, we proceed by introducing  the interdependence condition among the dual spaces of the subsystems.
First of all, we consider the injective map we introduced in equation \eqref{eqn: independence of preparation procedures} implementing the notion of independence among the states of the subsystems.
Then, we should implement the idea that, while a dual space possesses a ``linearity'' property, our ``composite'' objects are ``multilinear''.
Accordingly,  we assume the existence of an injective map
\be
\mathrm{I}^{*}\,\colon\,\stsp_{1}^{*}\,\times\,\cdots\,\times\stsp_{n}^{*}\,\lra\,\stsp^{*}
\ee
such that, introducing the notation
\be
f_{a_{1},...,a_{n}}\,:=\,\mathrm{I}^{*}(f_{a_{1}},\cdots,f_{a_{n}}),
\ee
we have
\be\label{eqn: compatibility condition for subsystems}
f_{a_{1},...,a_{n}}\left(\rho\right)\,=\, \prod_{j=1}^{n} \,f_{a_{j}}(\rho_{j})\,,
\ee
for every $\rho=\mathrm{I}(\rho_{1},\cdots,\rho_{n}) \in\mathrm{I}(\stsp_{1}\times\cdots\times\stsp_{n})\subset \stsp$.
We define elements of this type in $\stsp^{*}$ to be \grit{simple}.
The simple element $f_{a_{1},...,a_{n}}$ defined by equation \eqref{eqn: compatibility condition for subsystems} vanishes on (the injective image of) $\stsp_{1}\times\cdots\times\stsp_{n}$ whenever there is at least one $j\in[1,...,n]$ for which $f_{a_{j}}=0$.
This is in sharp contrast with what happens in the case $\stsp=\stsp_{1}\times\cdots\times\stsp_{n}$ (see equation \eqref{eqn: independent dual spaces}) where we need $f_{a_{j}}=0$ to be true for all $j\in[1,...,n]$ in order for the associated element in $\stsp^{*}$ to vanish on $\stsp_{1}\times\cdots\times\stsp_{n}$.
It is in this sense that we interpret equation \eqref{eqn: compatibility condition for subsystems} as an interdependence relation among the dual spaces of the subsystems.
There is a fully mature theory of non-commutative measure spaces called ``free probability theory'' (essentially, $C^*$-algebras with a tracial state), where it is introduced the notion of independence in a genuine noncommutative way and, what is more important, the notion of freeness (see \cite{voicolescu_dykema_nica-free_random_variables}). 
We believe that there is a connection between the notion of independence and freeness as defined in the context of free probability theory and the notions of independence and interdependence introduced above, however, we will analyse this connection elsewhere.

Now, we note  that  the existence of simple elements allows us to introduce the notion of \grit{separable} and \grit{entangled} states as follows.
First of all, consider the set $\stsp_{fs}$ composed by all those $\rho\in\stsp$  such that, for every simple element $f_{a_{1},...,a_{n}}\in\stsp^{*}$, there is a finite $N$, there are $n$-tuples $(\rho_{1}^{j},...,\rho^{j}_{n})$ with $j=1,...,N$ and $\rho_{k}^j$ in $\stsp_{k}$ for every $k\in[1,...,n]$, and there is a probability vector  $\vec{p}=(p_{1},...,p_{N})$ such that
\be\label{eqn: finite separable states}
\left(f_{a_{1},...,a_{n}}\right)(\rho)\,=\,\sum_{j=1}^{N}\,p_{j}\,\prod_{k=1}^{n}\,f_{a_{k}}(\rho_{k}^{j}) \, .
\ee
Elements in $\stsp_{fs}$ are called \grit{finitely-separable}.
Then, the space of separable elements $\stsp_{s}$ is given by the closure of $\stsp_{fs}$ in $\stsp$ with respect to a suitable topology that, in general, will depend on the specific situation considered.
For instance, if $\stsp$ is the space of states of a $C^{*}$ algebra $\appa$ (i.e., the space of positive, normalized linear functionals on $\appa$), the closure of $\stsp_{fs}$ in $\stsp$ is taken with respect to the weak* topology on $\stsp$ induced by $\appa$ when thought of as a subset of its double dual $\appa^{**}$.
It is not hard to see that the space of separable elements is a convex cone in $\stsp$.
An element in $\stsp$ which is not separable will be called \grit{entangled}, and, in general, composite systems admit entangled states.
In finite dimensions, classical probability theory is the only case in which there are no entangled states.

Below, we will show that the product convex set $\stsp =\stsp_{1}\times\cdots\times\stsp_{n}$ considered above is ruled out as a valid candidate because the interdependence condition among the dual spaces of the subsystems forces $\stsp^{*}$ to ``contain'' a copy of the tensor product $\otimes_{j=1}^{n}\,\stsp_{j}^{*}$ of the dual spaces of the single subsystems.
For this purpose, we define $W\subseteq \stsp^{*}$ to be the vector space obtained taking arbitrary but finite linear combinations of simple elements in $\stsp^{*}$, and we prove that $W$ is isomorphic, as a vector space, with the (algebraic) tensor product $\otimes_{j=1}^{n}\,\stsp_{j}^{*}$ by exploiting the universal property of the (algebraic) tensor product.
Essentially, we will see that, given any vector space $X$, and any multilinear map 
\be
\phi\,\colon\,\stsp_{1}^{*}\times\cdots\times\stsp_{n}^{*}\,\lra\, X,
\ee
there is a unique linear map
\be
\Phi\,\colon\,W\,\lra\,X
\ee 
such that 
\be\label{eqn: condition for tensor product}
\phi=\Phi\circ \mathrm{I}^{*}\,.
\ee
Recall that the range of $\mathrm{I}^{*}$ is in $W$ because it coincides with the set of simple elements generating $W$.
We start defining $\Phi$ on the simple elements in $W$ by setting 
\be
\Phi(f_{a_{1},...,a_{n}})\,:=\,\phi(f_{a_{1}}^{1},\cdots,f_{a_{n}}^{n})\,.
\ee
Since the set of simple elements is a generating set for $W$, we can extend $\Phi$ to the whole $W$ by linearity so that, by construction, we have that equation \eqref{eqn: condition for tensor product} holds.
Furthermore, again because the set of simple elements is a generating set for $W$, the map $\Phi$ is unique by construction.
Consequently, the universal property of the algebraic tensor product implies the existence of a vector space isomorphism between $W$ and $\otimes_{j=1}^{n}\,\stsp_{j}^{*}$.
It is important to note that, in general, $W$ is only a proper subspace of $\stsp^{*}$.

Now, it is not hard to see that a convex set $\stsp$  generating a vector space $\V$ which is isomorphic with the the tensor product $\otimes_{j=1}^{n}\V_{j}$ of the vector spaces generated by the single $\stsp_{j}$'s may always be interpreted as the convex set of a composite system implementing the independence condition among states of the subsystems (see equation \eqref{eqn: independence of preparation procedures}) and with the interdependence condition among the dual spaces of the subsystems (see equation \eqref{eqn: compatibility condition for subsystems}).
Indeed, we can define the map $\mathrm{I}\colon \stsp_{1}\times\cdots\times\stsp_{n}\,\lra\,\stsp$ given by
\be
(\rho_{1},\cdots,\rho_{n})\,\mapsto\,\mathrm{I}(\rho_{1},\cdots,\rho_{n})\,=\,\rho_{1}\,\otimes\,\cdots\,\otimes\,\rho_{n},
\ee 
and a general result from  linear algebra assures us that $\otimes_{j=1}^{n}\,\stsp_{j}^{*}$ is always a subset of $\stsp^{*}$ (recall that we defined the dual space of a convex set to be the dual space of the vector space generated by the convex set itself).
Furthermore, in the finite-dimensional case where $\mathrm{dim}(\V_{j})<\infty$ for all $j\in[1,...,n]$, we have that 
\be
W\cong\otimes_{j=1}^{n}\,\stsp^{*}_{j}\,\cong\,\left(\otimes_{j=1}^{n}\V_{j}\right)^{*},
\ee
where $\V_{j}$ is the vector space generated by $\stsp_{j}$.
Consequently, choosing the vector space $\V$ generated by $\stsp$ to be the tensor product $\otimes_{j=1}^{n}\V_{j}$ is equivalent to impose the minimality condition $\stsp^{*}=W$ for the dual space of $\stsp$.
Note that this is no longer true in the infinite-dimensional case because the dual space of a tensor product need not be the tensor product of the dual spaces.
However, it is reasonable to say that the subleties associated with infinite dimensions requires more structures to be handled, and the framework of arbitrary convex spaces is too broad to provide these structures.

As a final comment, let us point out that, even if we consider the finite-dimensional case with the minimality condition $\stsp^{*}=W$, there is no way to single out unambigously an explicit candidate for $\stsp$ without introducing further assumptions.
Again, this should not come as a surprise because the theoretical framework of arbitrary convex spaces considered here is too broad.

\section{Conclusions}

We investigated the notion of composite system made of distinguishable parties in the context of physical theories for which the admissible spaces of states are real, convex spaces.
Essentially, we modelled a composite system by means of the family $\{\stsp_{j}\}_{j\in[1,...n]}$ of spaces of states of the $n$ subsystems together with the space $\stsp$ of states of the total system ``endowed'' with two mathematical constraints.
First of all, we imposed an independence relation among the states of the subsystems in terms of an injective linear map $\mathrm{I}\colon \stsp_{1}\times\cdots\times\stsp_{n}\,\lra\,\stsp$, where $\stsp$ is the space of states of the total system, and $\stsp_{1}\times\cdots\times\stsp_{n}$ is the Cartesian product of the spaces of states of the subsystems.
From the operational point of view, the existence of the map $\mathrm{I}$ should be thought of as guaranteeing that each party of the system is free to prepare its associated subsystem in any of the possible states independently from the preparations of the other parties.
Then, we introduced an interdependence condition among the dual spaces of the subsystems (see equation \eqref{eqn: compatibility condition for subsystems}).
We saw that these two mathematical conditions are  enough to  introduce the notion of separable and entangled states in the context of arbitrary convex spaces, and to prove that the dual space $\stsp^{*}$ of a composite system must contain a copy of the tensor product $\otimes_{j=1}^{n}\stsp_{j}^{*}$ of the dual spaces of the single subsystems.
Furthermore,  in the finite-dimensional case, $\stsp$ generates a vector space $\V$ which is isomorphic with the tensor product $\otimes_{j=1}^{n}\V_{j}$ of the vector spaces generated by the single subsystems if and only if $\stsp^{*}$ satisfies a minimality condition.

We must stress that the interdependence condition expressed by equation \eqref{eqn: compatibility condition for subsystems} has not yet a clear physical interpretation, but its mathematical expression points toward a connection with the notions of independence and freeness as defined in the context of free probability theory (see \cite{voicolescu_dykema_nica-free_random_variables}) which will be explored elsewhere.

\section*{Acknowledgements}

A.I. and G.M. acknowledge financial support from the Spanish Ministry of Economy and Competitiveness, through the Severo Ochoa Programme for Centres of Excellence in RD (SEV-2015/0554).
A.I. would like to thank partial support provided by the MINECO research project MTM2017-84098-P and QUITEMAD++, S2018/TCS-­4342.
G.M. would like to thank the support provided by the Santander/UC3M Excellence Chair Programme 2019/2020.

%\addcontentsline{toc}{section}{References}
%\bibliographystyle{plain}
%\bibliography{scientific_bibliography}

\end{document}